\input{aipcheck}

\documentclass[
    ,final            % use final for the camera ready runs
  ]
  {aipproc}

\newcommand{\bb}{\begin{thebibliography}}
\newcommand{\eb}{\end{thebibliography}}

\newcommand{\bea}{\begin{eqnarray}}
\newcommand{\ba}{\begin{array}}
\newcommand{\ea}{\end{array}}
\newcommand{\eea}{\end{eqnarray}}

\newcommand{\Tr}{{\rm Tr}}

\newcommand{\N}{{\cal N}}

\layoutstyle{6x9}

\begin{document}

\title{Novel Transversity Properties in SIDIS}

\author{Leonard Gamberg}{
  address={Division of Science,
Penn State-Berks Lehigh Valley College, Reading, PA 19610,
USA}
}

\author{Gary R. Goldstein}{
  address={Department of Physics
          and Astronomy, Tufts University,
           Medford, MA 02155, USA}
}

\author{Karo A. Oganessyan}{
 address={INFN-Laboratori Nazionali di Frascati, Enrico Fermi 40,
  I-00044 Frascati, Italy}
  ,altaddress={DESY, Notkestrasse 85, 22603 Hamburg, Germany}
}

\begin{abstract}
We consider a  rescattering mechanism to calculate a leading 
twist $T$-odd pion  
fragmentation function, a candidate for filtering the transversity 
properties of the nucleon. We evaluate the single spin azimuthal asymmetry  
for a transversely polarized target in semi-inclusive deep inelastic 
scattering (for HERMES kinematics) and  the double  
$T$-odd $\cos2\phi$ asymmetry in this framework.
\end{abstract}

\maketitle

%%%%%%%%%%%%%%%%%%%%%%%%%%%%%%%%%%%%%%%%%%%%
%% MAINMATTER
%%%%%%%%%%%%%%%%%%%%%%%%%%%%%%%%%%%%%%%%%%%%
\subsubsection{Introduction}
\vskip-0.35cm
The transversity distribution, $h_1$  which 
measures the probability to find a transversely polarized quark in the 
transversely polarized nucleon, is as important for the description 
of the internal nucleon spin structure as the more familiar 
helicity distribution function, $g_1$. However, it still remains 
unmeasured, unlike the spin-average and helicity distribution 
functions, which are known experimentally and extensively modeled 
theoretically. The difficulty is that $h_1$ is chiral odd, 
and consequently suppressed in 
inclusive deep inelastic scattering (DIS) processes~\cite{jaffe91}; it has 
to be accompanied by a second chiral-odd quantity.  
Semi-inclusive deep inelastic scattering (SIDIS) on
polarized nucleons is one of 
several~\cite{SMC,HERMES,STAR} promising methods  proposed to access
transversity. It relies on just such a quantity
the so called Collins fragmentation function~\cite{cnpb93}, which  
correlates the transverse spin of the fragmenting quark to the transverse 
momentum of the produced hadron. Beside being chiral-odd, this fragmentation 
function is also time-reversal odd ($T$-odd)~\cite{boer,ANSL} 
which makes its 
calculation challenging.  
In this context, the  non-zero single spin asymmetries in  recent  
measurements~\cite{HERMES} may
signal the existence of a non-trivial $T$-odd effects which are
intimately tied to our understanding of transversity.
Here we  explore~\cite{gambH} a  one-gluon exchange 
mechanism, for the fragmentation 
of a transversely polarized quark into a spinless
hadron  similar to the approach we 
applied~\cite{gamb_gold_ogan1,gamb_gold_ogan2} 
to the distribution of the transversely polarized quarks in 
the both unpolarized and transversely polarized nucleons. 
The non-perturbative information about the quark content of the target 
and the fragmentation of quarks into hadrons 
in SIDIS is encoded in 
the general form of the factorized cross sections in terms of  
the quark distributions $\Phi(p)$ and  fragmentation functions $\Delta(k)$, 
 entering the hadronic tensor.  
To leading 
order in $1/Q^2$~\cite{mulders2} the fragmentation functions are projected from
\bea
\Delta(k,P_h) \hskip -.10cm=\hskip -.10cm
\sum_X \int 
\frac{d\xi^+d^2 \xi_\perp}{2z\,(2\pi)^3} \ 
e^{ik\cdot \xi} \,\langle 0\ 
\vert 
{\cal G}_{{\scriptscriptstyle{[\xi^+,-\infty]} } } 
\psi (\xi) 
\vert X;P_h\rangle\langle X;P_h\vert \overline \psi(0) 
{\cal G}^\dagger_{{\scriptscriptstyle{[0,-\infty]}}}
\vert 0 \rangle \Big|_{\xi^- = 0}.
\label{FF} 
\eea
Here $k$  quark 
fragmenting momenta and $P_h$ is the
fragmented hadron momenta. 
The path ordered exponential along the light like 
direction $\xi^-$ is 
$${\cal G}_{[\xi^-,\infty]}={\cal P}
\exp{\left(-ig\int_{\xi^-}^\infty d\xi^- A^+(\xi)\right)}.
$$
In non-singular gauges~\cite{cplb,ji}, the gauge link   gives rise
to initial and final  state interactions which in turn 
provide a mechanism to generate leading  twist $T$-odd 
contributions to both the distribution and {\em fragmentation}
functions.  
The joint product of these functions enter
novel azimuthal asymmetries and single spin asymmetries (SSAs) 
that have been reported in the 
literature~\cite{bhs,gold_gamb,gamb_gold_ogan1,gamb_gold_ogan2}.
 Such an analysis was recently applied to the $T$-odd 
$f_{1T}^\perp$~\cite{bhs,ji,sivers} 
and $h_1^\perp$~\cite{gold_gamb,gamb_gold_ogan1,gamb_gold_ogan2} 
distribution functions in addition to 
$T$-odd baryon fragmentation functions~\cite{metz}.
We apply~\cite{gambH} an analogous procedure to generate 
the $T$-odd pion fragmentation 
function, $H_1^\perp(z)$ (see also~\cite{bac_metz}). 
\subsubsection{Pion Fragmentation Function}
\begin{figure}
\includegraphics[height=4.0cm]{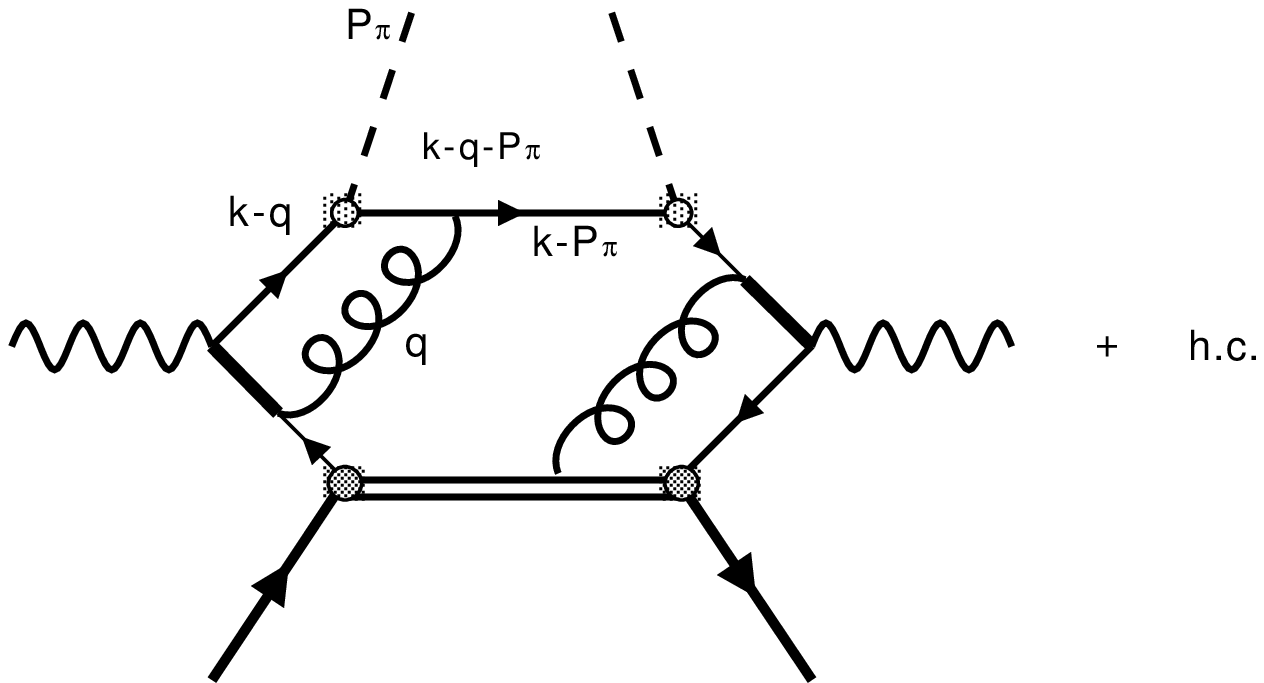}
\includegraphics[width=5.0cm,height=4.0 cm]{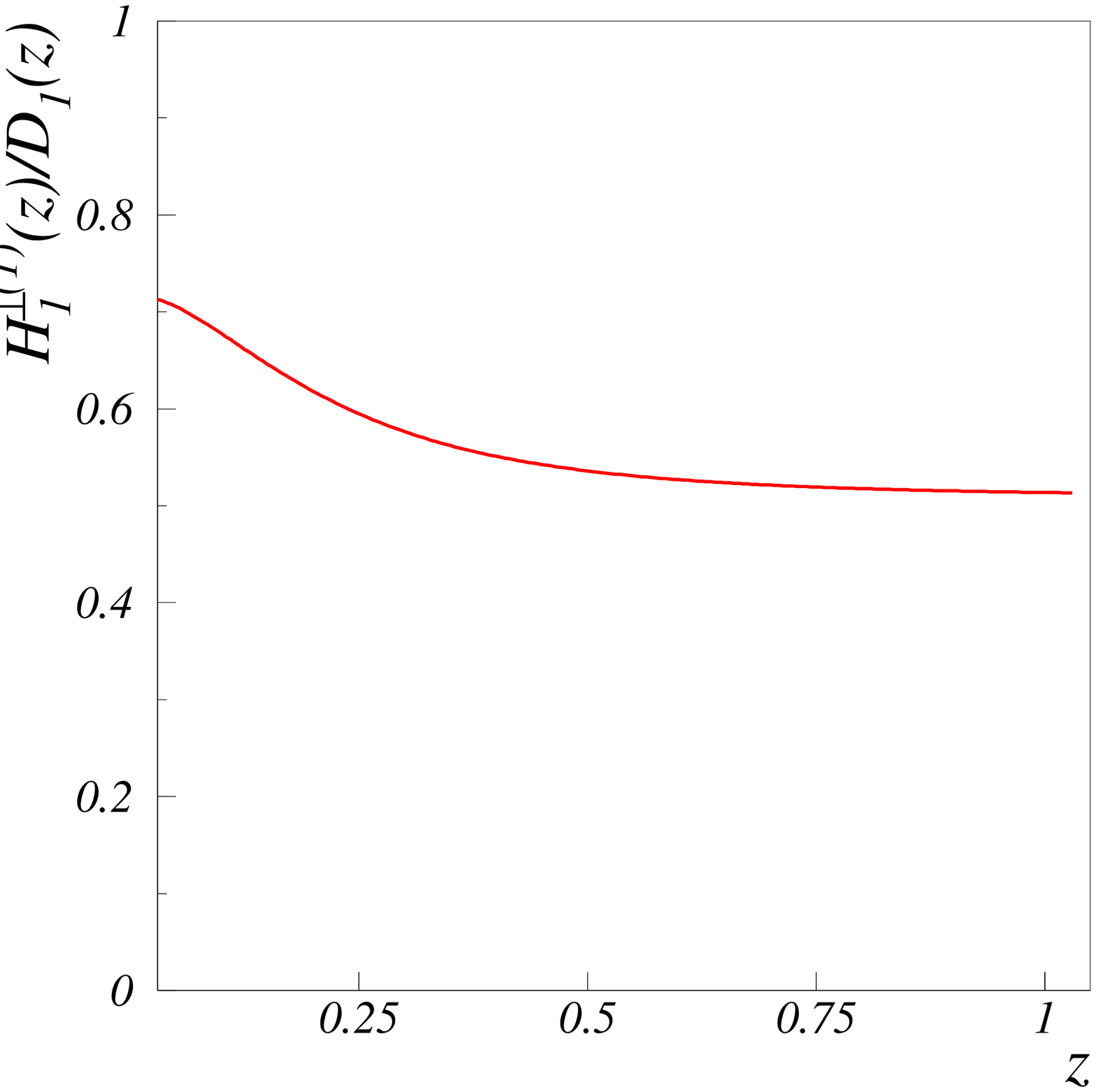}
\caption{\label{analyze} 
Left Panel: $h_1^\perp\star H_1^\perp$ $\cos 2\phi$ 
asymmetry. 
Right Panel: 
The weighted analyzing power
$H_1^{\perp (1)}(z)/D_1(z)$ as a function of $z$.}
\end{figure}

\vskip -.35cm
The leading order  contributions 
to the $T$-odd  fragmentation functions 
come from the first non-trivial term in expanding
 the path ordered gauge link operator.
The corresponding Feynman rules are those 
for interactions between an eikonalized struck
quark and the remaining target~\cite{col82} depicted in Fig.~\ref{analyze}.  
In modeling the highly off-shell
fragmenting quark we adopt a minimal spectator~\cite{hood} approach.
We couple the on-shell spectator, as a quark  
interacting with the produced pion 
through a Gaussian distribution
in the transverse momentum dependence of the quark-spectator-pion 
vertex ~\cite{gamb_gold_ogan2,gambH} in order to address
the $\log$ divergence
arising in  the moments of fragmentation functions. 
The leading order (in $1/Q$) one loop contribution 
which arises in the limit that the
virtual photon's
momentum becomes large
corresponding to the rescattering
of the initial state quark depicted  in Fig.~\ref{analyze}.
The resulting  twist 2, $T$-odd  
contribution  the  fragmentation function~\cite{gambH}
projected from $\Tr\left(\gamma^-\gamma^\perp\gamma_5\Delta\right)$ is
$$H_1^\perp(z,k_\perp) 
\hspace{-.10cm}=\hspace{-.10cm}
\frac{{\N^\prime}^2 f^2g^2}{(2\pi)^4}\frac{1}{4z}\frac{(1-z)}{z}
\frac{m}{\Lambda^\prime(k^2_\perp)}
\frac{M_\pi}{k_\perp^2}
e^{{\scriptstyle{-b^\prime\left(k^2_\perp- \Lambda^\prime(0)\right)}}}
\left[\Gamma(0,{\scriptstyle{b\Lambda^\prime(0)}})
\hspace{-.10cm}-\hspace{-.10cm}
\Gamma(0,{\scriptstyle{b^\prime\Lambda^\prime(k^2_\perp)}})\right] ,
$$
$\Gamma(0,z)$ is the incomplete gamma function and 
$\Lambda^\prime(k^2_\perp)=k_\perp^2 +\frac{1-z}{z^2}M_\pi^2+ 
\frac{\mu^2}{z} -\frac{1-z}{z} m^2$. 
The average $<k^2_{\perp}>=1/b^\prime$ is a regulating scale 
which we fit to the expression for the integrated unpolarized 
fragmentation function 
$$
D_1(z)=\frac{{\N^\prime}^2 f^2}{4(2\pi)^2}
\frac{1}{z}\frac{\left(1-z\right)}{z}
\{\frac{m^2-\Lambda^\prime(0)}{\Lambda^\prime(0)}
-\left[2b^\prime\left(m^2-\Lambda^\prime(0)\right)-1\right]
e^{2b^\prime\Lambda^\prime(0)}
\Gamma(0,{\scriptstyle{2b^\prime\Lambda^\prime(0)}})\}
$$
which is in  good agreement with the
distribution of Ref.~\cite{kretz}.
In Fig.~\ref{analyze} the weighted 
the analyzing power, $H_1^{\perp (1)}(z)/ D_1(z)$, is displayed. 
The resulting behavior is similar to a previous model ansatz 
proposed by Collins and calculated in Ref.~\cite{kotz}.
The $\cos2\phi$ asymmetry of SIDIS is projected out of the cross section 
and depends on a leading double $T$-odd product, 
\bea
\langle \frac{\vert P^2_{h{\perp}} \vert}{M M_\pi} \cos2\phi \rangle
{\scriptscriptstyle_{UU}}&=& 
\frac{{8(1-y)} \sum_q e^2_q h^{\perp(1)}_1(x) z^2 H^{\perp(1)}_1(z)}
{{(1+{(1-y)}^2)}  \sum_q e^2_q f_1(x) D_1(z)}.
\label{ASY_cos} 
\eea
$UU$ indicates unpolarized beam and 
target  and $h_1^{\perp (1)}(z)$ is the weighted
moment of the distribution  
function~\cite{boer,gamb_gold_ogan2}.
For a transversely polarized target 
nucleon, the $\sin(\phi+\phi_s)$ asymmetry\cite{cnpb93,mulders2} 
can be similarly obtained yielding, the convolution of two chiral-odd 
structures,
\bea
\langle \frac{P_{h\perp}}{M_\pi}
\sin(\phi+\phi_s) \rangle_{\scriptscriptstyle UT}
&=& 
\big|S_T\big|\frac{2(1-y) \sum_q e^2_q h_1(x) z H^{\perp(1)}_1(z)}
{{(1+{(1-y)}^2)}  \sum_q e^2_q f_1(x) D_1(z)}.
\label{ASY_sin}
\eea
\begin{figure}
\includegraphics[height=4.0 cm,width=3.7cm]{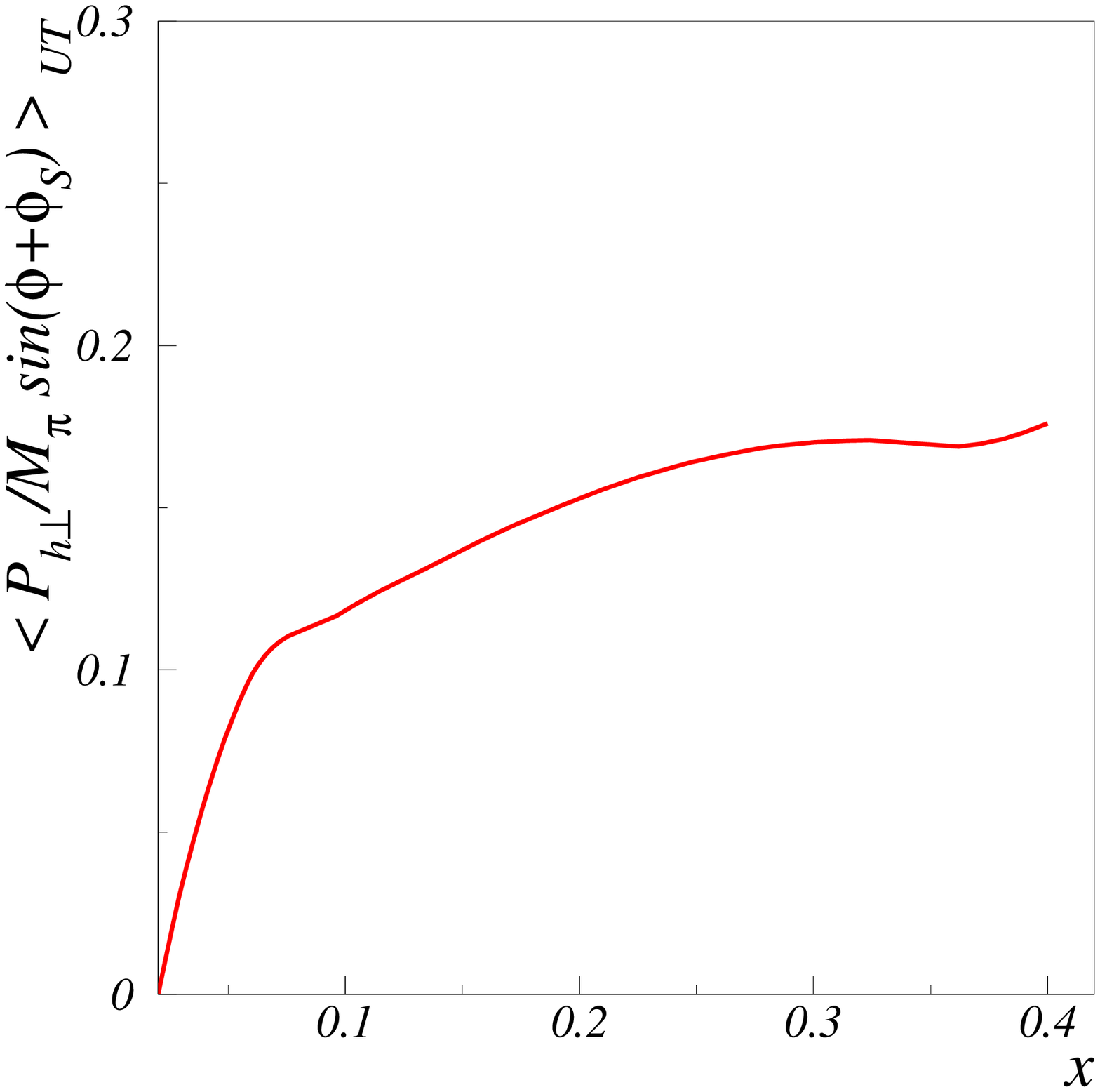}
\includegraphics[height=4.0 cm,width=3.7cm]{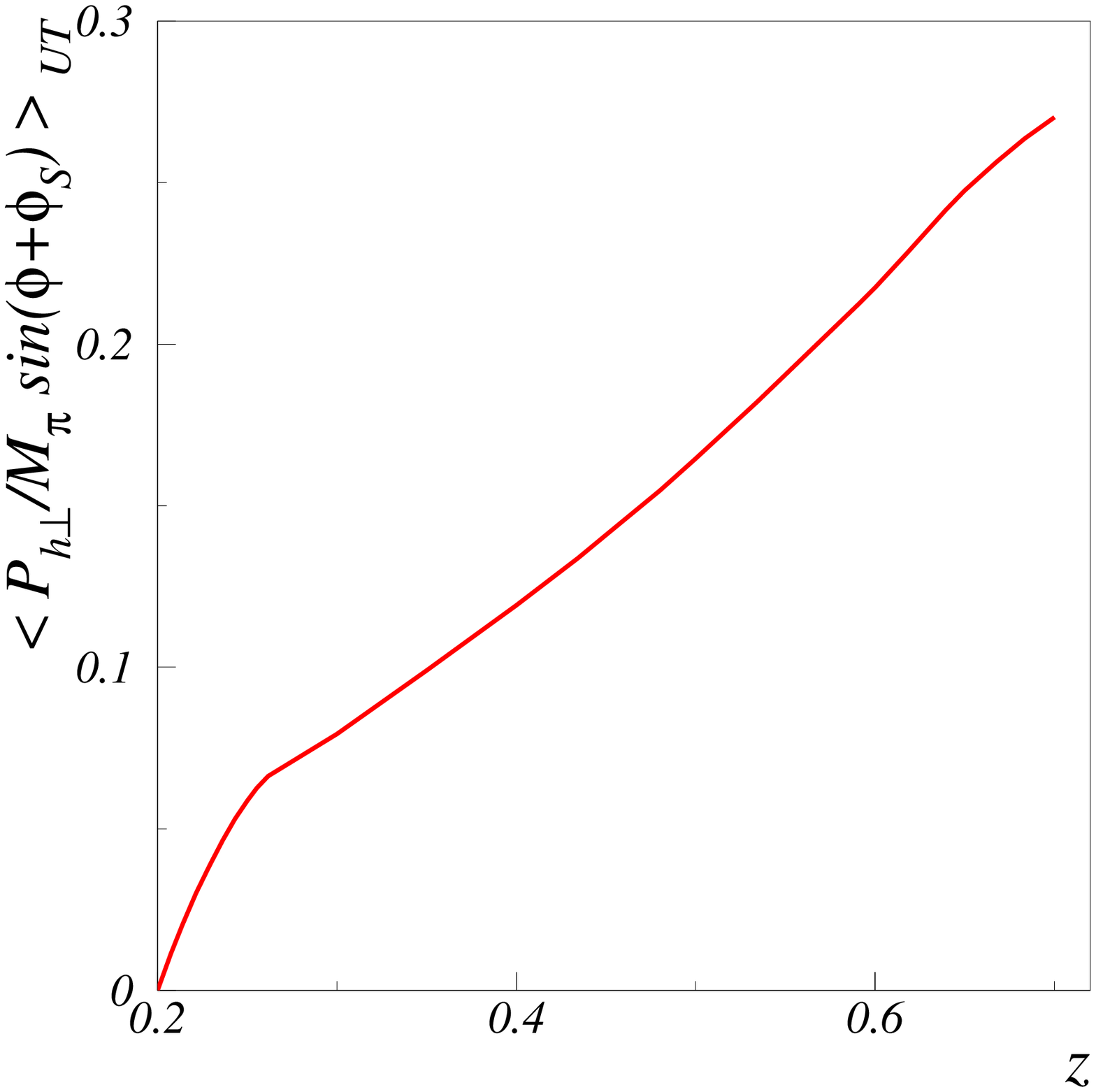}
\includegraphics[height=4.0 cm,width=3.7cm]{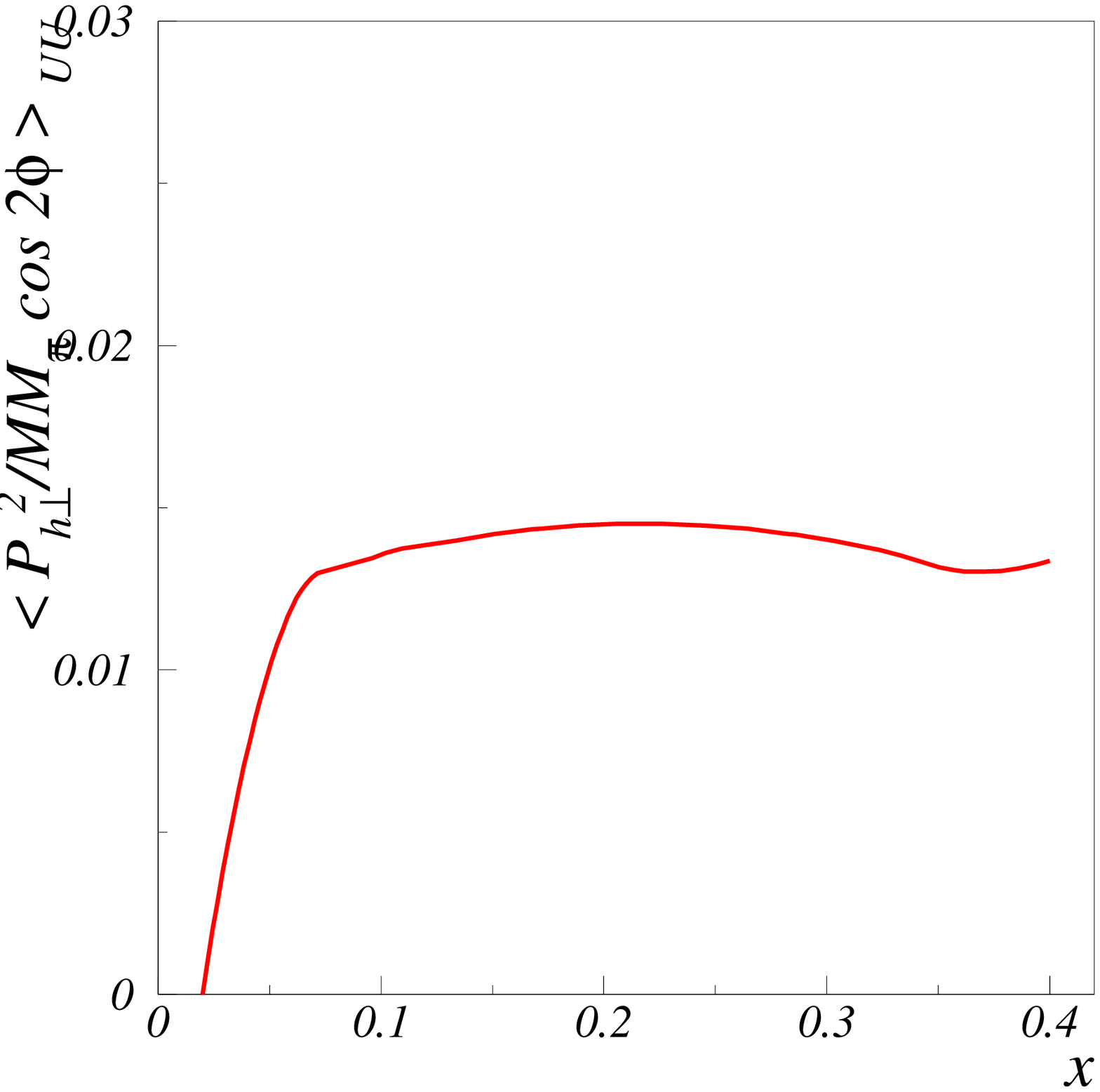}
\includegraphics[height=4.0 cm,width=3.7cm]{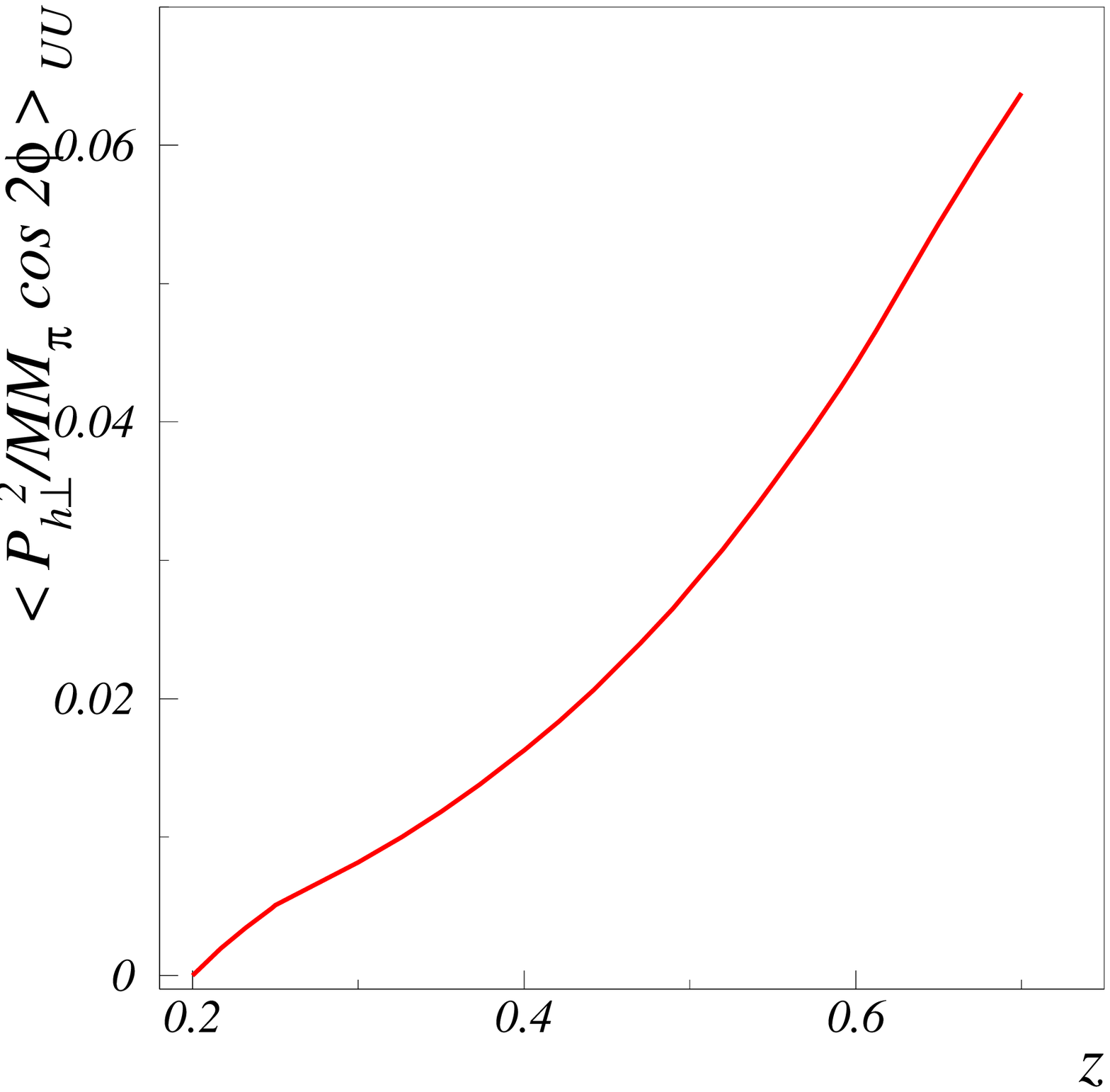}
\caption{\label{collins}   The 
\protect{${\langle \sin(\phi+\phi_s) 
\rangle}_{\scriptscriptstyle UT}$ } asymmetry for $\pi^+$ production 
as  a function of  $x$ and $z$. 
The \protect{${\langle \cos2 \phi \rangle}_{\scriptscriptstyle UU}$} 
asymmetry for \protect{$\pi^+$} production 
as a function of \protect{$x$} and  \protect{$z$}.}
\end{figure}
The variable range to coincides with the HERMES 
kinematics~\cite{gambH}.  In Fig.~\ref{collins} the 
asymmetry of Eq.~(\ref{ASY_sin}) 
for $\pi^+$ production on a proton target is presented 
as a function of $x$ and $z$, respectively 
indicating approximately a  
$10-15\%$  $P_{h\perp}/M_{\pi}$ weighted $\sin(\phi+\phi_s)$  asymmetry.
Also, in Fig.~\ref{ASY_sin} the  $P^2_{h\perp}/(M M_{\pi})$ weighted 
$\cos 2\phi$ asymmetry of Eqs.~(\ref{ASY_cos}) 
for $\pi^+$ production on an unpolarized proton target is presented as 
a function of $x$ and $z$, respectively indicating 
a few percent effect.  
\vskip-1cm
\subsubsection{Conclusion}
\vskip-0.35cm
A mechanism to generate the $T$-odd Collins 
fragmentation function that is derived from the gauge link
has been considered. This approach complements 
 the approach that was employed to 
generate the $T$-odd distribution functions,    $f_{1T}^\perp$ and $h_1^\perp$
that fuel the Sivers and $\cos 2\phi$ asymmetries. 
The derivation of $H_1^\perp$ 
is consistent with the observation that intrinsic transverse quark momenta and 
angular momentum conservation are intimately tied with studies of 
transversity. 
Furthermore, this approach is interesting in that it does not suffer from
the possible cancellation of the Collins effect cited in~\cite{jin_jaffe}.
This effect is generated in the 
non-trivial phase associated with the gauge link 
operator~\cite{cplb,ji,metz,gamb_gold_ogan1,gamb_gold_ogan2,pij}.
We have evaluated the
analyzing power and predicted
the $P_{h\perp}/M_{\pi}$ weighted $\sin(\phi+\phi_S)$ asymmetry
at HERMES energies. Additionally, we 
predict that  there is a non-trivial $\cos2\phi$ asymmetry associated with the
asymmetric distributions of transversely polarized quarks inside
unpolarized hadrons. 
Generalizing from these model calculations, it is clear that initial and 
final state  interactions can account for  leading twist 
$T$-odd contributions to SSAs.  
Using rescattering as a mechanism to generate 
$T$-odd distribution and fragmentation functions 
opens a new window into the theory and phenomenology of transversity in 
hard processes.
\vspace*{-0.75cm}

\subsubsection{Acknowledgments}
\vskip-0.4cm
{\footnotesize
L.G. thanks the organizers of CIPANP for the invitation to present
this work. 
%%%%%%%%%%%%%%%%%%%%%%%%%%%%%%%%%%%%%%%%%%%%%%%%
%% You may have to change the BibTeX style below, depending on your
%% setup or preferences.
%%
%% If the bibliography is produced without BibTeX comment out the
%% following lines and see the aipguide.pdf for further information.
%%
%% For The AIP proceedings layouts use 
%\bibliographystyle{aipproc}   % if natbib is available
%\bibliographystyle{aipprocl} % if natbib is missing
%%%%%%%%%%%%%%%%%%%%%%%%%%%%%%%%%%%%%%%%%%%
%% You probably want to use your own bibtex database here
%%%%%%%%%%%%%%%%%%%%%%%%%%%%%%%%%%%%%%%%%%%
%\vskip-0.25
\vspace{-0.5cm}
\bb{99}
\bibitem{jaffe91} R. L. Jaffe and 
X. Ji, Phys. Rev. Lett. {\bf 67}, 552 (1991) .

\bibitem{SMC}  A. Bravar (Spin Muon Collaboration), Nucl. Phys. Proc. 
Suppl., {\bf 79} 520 (1999). 

\bibitem{HERMES} A. Airapetian {\it et al.}, Phys. Rev. Lett.
 {\bf 84}, 4047 (2000); Phys. Lett. B {\bf 562}, 182 (2003).

\bibitem{STAR} L. C. Bland, hep-ex/0212013; G. Rakness, hep-ex/0211068.  

\bibitem{cnpb93} J.C. Collins, Nucl. Phys. {\bf B396},161 (1993).

\bibitem{boer} D. Boer and P. J. Mulders, Phys. Rev. D {\bf 57}, 5780 (1998).

\bibitem{ANSL} M. Anselmino and F. Murgia, Phys. Lett B {\bf 442}, 470 (1998).

\bibitem{gambH} L. P. Gamberg, G. R. Goldstein and 
K.A.~Oganessyan, hep-ph/0307139, {\em To appear in Phys. Rev. D}.

\bibitem{gamb_gold_ogan1} L. P. Gamberg, G. R. Goldstein and 
K.A.~Oganessyan, hep-ph/0211155, {\em Proceedings
of the $15^{\rm th}$ International Spin Physics Symposium (SPIN 2002)}, 
Long Island, New York, September 2002. 

\bibitem{gamb_gold_ogan2} L. P. Gamberg, G. R. Goldstein and 
K.A.~Oganessyan, Phys. Rev. D {\bf 67 }, 071504 (2003).

\bibitem{mulders2} R. D. Tangerman and P. J. Mulders,
Phys. Lett. B {\bf 352}, 129 (1995); Nucl. Phys. {\bf B461}, 197 (1996).

\bibitem{cplb} J. C. Collins, Phys. Lett. B {\bf 536}, 43 (2002).

\bibitem{ji} X. Ji and F. Yuan, Phys. Lett. B {\bf 543}, 66 (2002);
A.V. Belitsky, X. Ji and F. Yuan, Nucl. Phys. B {\bf 656}, 156 (2003) .

\bibitem{bhs} S. Brodsky, D.S. Hwang and I. Schmidt, 
Phys. Lett. B {\bf 530}, 99 (2002); D. Boer, S. Brodsky, D.S. Hwang, 
Phys. Rev D {\bf 67}, 054003  (2003).

\bibitem{gold_gamb} G. R. Goldstein and L. P. Gamberg, hep-ph/0209085,
{\em Proceedings of $31^{\rm st}$ International Conference on 
High Energy Physics (ICHEP 2002)}, Amsterdam, The Netherlands, Jul 2002.

\bibitem{sivers} D. Sivers, 
Phys. Rev D 41 (1990) 83 ; Phys. Rev. D {\bf 43}, 261 
(1991).

\bibitem{metz} A. Metz, Phys. Lett. B {\bf 549}, 139 (2002).

\bibitem{bac_metz} A. Bacchetta, A. Metz, and J. J. Yang,
hep-ph/0307282.

\bibitem{col82} J. C. Collins and D. E, Soper, Nucl. Phys. {\bf B194},
445 (1982).

\bibitem{hood} P. Hoodbhoy, Phys.Rev. D {\bf 51}, 32 (1995).

\bibitem{kretz} S. Kretzer, Phys. Rev. D {\bf 62}, 054001 (2000).

\bibitem{kotz} A. M. Kotzinian and P. J. Mulders,
Phys. Lett. B {\bf 406}, 373 (1997).

\bibitem{jin_jaffe} R. L. Jaffe, X. Jin, and J. Tang, 
Phys. Rev. Lett {\bf 80}, 1166 (1998).

\bibitem{pij} D.  Boer, P.J. Mulders, and F. Pijlman,  hep-ph/0303034.

\eb

%%%%%%%%%%%%%%%%%%%%%%%%%%%%%%%%%%%%%%%%%%%
%% Just a reminder that you may have to run bibtex
%% All of it up to \end{document} can be removed
%% if you don't like the warning.
%%%%%%%%%%%%%%%%%%%%%%%%%%%%%%%%%%%%%%%%%%%
\IfFileExists{\jobname.bbl}{}
 {\typeout{}
  \typeout{******************************************}
  \typeout{** Please run "bibtex \jobname" to optain}
  \typeout{** the bibliography and then re-run LaTeX}
  \typeout{** twice to fix the references!}
  \typeout{******************************************}
  \typeout{}
 }

\end{document}